\documentclass[fleqn,usenatbib]{mnras}

\usepackage{newtxtext,newtxmath}

\usepackage[T1]{fontenc}

\DeclareRobustCommand{\VAN}[3]{#2}
\let\VANthebibliography\thebibliography
\def\thebibliography{\DeclareRobustCommand{\VAN}[3]{##3}\VANthebibliography}

\usepackage{graphicx}	
\usepackage{amsmath}	

\title[BBH Mergers in AGNs vs. NSCs]{Binary Black Hole Merger Rates in AGN Disks versus Nuclear Star Clusters: Loud beats Quiet}

\author[K. E. S. Ford et al.]{
K. E. Saavik Ford,$^{1,2,3,4}$\thanks{E-mail: sford@amnh.org (KESF)}
and Barry McKernan,$^{1,2,3,4}$
\\
$^{1}$Department of Astrophysics, American Museum of Natural History, New York, NY 10024, USA\\
$^{2}$Center for Computational Astrophysics, Flatiron Institute, New York, NY 10010, USA\\
$^{3}$Graduate Center, City University of New York, 365 5th Avenue, New York, NY 10016, USA\\
$^{4}$Department of Science, BMCC, City University of New York, New York, NY 10007, USA\\
}

\date{Accepted XXX. Received YYY; in original form ZZZ}

\pubyear{2015}

\begin{document}
\label{firstpage}
\pagerange{\pageref{firstpage}--\pageref{lastpage}}
\maketitle

\begin{abstract}
Galactic nuclei are promising sites for stellar origin black hole (BH) mergers, as part of merger hierarchies in deep potential wells. 
We show that binary black hole (BBH) merger rates in active galactic nuclei (AGN) should always exceed merger rates in quiescent galactic nuclei (nuclear star clusters, NSCs) around supermassive BHs (SMBHs) without accretion disks. This is primarily due to average binary lifetimes in AGN that are significantly shorter than in NSCs. The lifetime difference comes from rapid hardening of BBHs in AGN, such that their semi-major axes are smaller than the hard-soft boundary of their parent NSC; this contrasts with the large average lifetime to merger for BBHs in NSCs around SMBHs, due to binary ionization mechanisms. Secondarily, merger rates in AGNs are enhanced by gas-driven binary formation mechanisms. Formation of new BHs in AGN disks are a minor contributor to the rate differences. With the gravitational wave detection of several BBHs with at least one progenitor in the upper mass gap, and signatures of dynamical formation channels in the $\chi_{\rm eff}$ distribution, we argue that AGN could contribute $\sim 25\%-80\%$ of the LIGO-Virgo measured rate of $\sim 24 \rm{Gpc}^{-3} \rm{yr}^{-1}$.
\end{abstract}




\section{Introduction}

Many binary black hole (BBH) mergers have now been observed, but there is not yet sufficient evidence to disentangle the relative amplitude of contributions from various proposed merger channels \citep{O3a}. Given that the observed $\chi_{\rm eff}$ distribution possesses a non-negligible negative component, it seems likely that dynamical channels play a significant role in BBH mergers observed to date\footnote{While isolated binary evolution can produce mergers with negative $\chi_{eff}$, the black hole natal supernova kick velocities would then be required to be $O(10^3)$km/s \citep{Brandt95}, a requirement in tension with observed distributions of velocities of galactic HMXBs \citep{Repetto17}}. 

Most dynamical BBH merger channels are characterized by a high expected number density of black holes in that environment. In particular, BBH mergers are expected in globular clusters (GCs) \citep{Rodriquez16a} and in nuclear star clusters (NSCs) \citep{OLeary09,Antonini14,Fragione19BHNS}. Proposed dynamical merger sites with lower BH number density include more numerous open clusters \citep[e.g.][]{Mapelli20}. Importantly, AGN are a dynamical BBH merger channel   \citep[e.g.][]{McK12,McK14,Bellovary16,Bartos17,Stone17,McK18,Secunda20a,Tagawa20} which generate parameters distinguishable from other dynamical channels. Indeed, a possible anti-correlation between effective spin and mass ratio among LIGO-Virgo gravitational wave (GW) detected BBH mergers \citep{Tom2021} might be a signature of the AGN channel \citep{qXeff21}, but at present it remains challenging to identify contributions from the different dynamical merger channels.

Hierarchical mergers are especially useful discriminants, both between non-dynamical and dynamical channels, as well as among different dynamical channels. Hierarchical mergers are expected from dynamics, as long as merger products are retained in the same environment, and form an identifiable population based on their GW measured parameters alone--at least one progenitor mass component in the upper mass gap, and with high spin as a result of a prior merger. Every individual hierarchical merger detected ($ng-mg, n>1,m \geq 1$ where $g$ denotes the merger `generation' of a progenitor BH) constrains the general contribution from dynamics, since each $ng-mg$ merger requires multiple ($1g-1g$) mergers \textit{from the same channel}\footnote{Also noted by \cite{Gerosa19}}. Consequently, predictions of rate ratios ${\cal{R}}_{ng-mg}/{\cal{R}}_{1g-1g}$ with $n>1, m \geq 1$, from dynamical channels allow us to identify the likely fraction of $1g-1g$ mergers (though not the individual events) from the different dynamical channels. Since different channels produce different expected mass and and spin distributions, if we can firmly identify a small number of events as uniquely attributable to a single channel, we can also hope to disentangle the `mixing fraction' between channels, and their parameter distributions by subtracting off the contribution of a single well-identified channel.

Dynamical BBH mergers can occur in shallow potential wells (e.g. GCs or open clusters) or in deep potential wells (e.g. NSCs or AGN). For BH with non-negligible natal spins, the kick velocity ($v_{\rm k}$) generated by a $1g-1g$ BBH merger is expected to be $v_{k}>50{\rm km/s}$, i.e. the escape velocity of present-day globular clusters \citep{Gerosa19}. However, recent work \citep{Carl21} suggests that the \textit{local} escape velocity at the location of the most massive mergers in super-massive GCs (up to $10^{8}M_{\odot}$) can be higher, up to $v_{\rm esc}>120{\rm km/s}$. So, hierarchical mergers up to $3g-3g$ can occur from the most over-massive globular clusters \emph{if and only if} the natal spins of BH are extremely small \citep[e.g.][]{FullerMa19}. However, the highest generation of mergers \textit{due to} GCs always occur after the bound BBH is ejected from the cluster \citep{Carl21}. This means that the highest generation of BBH mergers originating in GCs should always have circularized by the time they reach the frequencies of ground-based GW detectors. The rate  of mergers of later BH generations ($n>3$) originating from GCs are always strongly suppressed due to the relatively low escape velocity of GCs (even accounting for their mass evolution over cosmic time). So, the detection of a $ng-mg$ mergers with $n>3, m \geq 1$ would strongly suggest a merger origin in a deep potential well \textit{along with} an additional substantial fraction of the GW detected lower generation ($ng-mg, n\leq 3$) mergers from the same origin. Even detection of a $3g-mg$ merger suggests a dynamical origin unrelated to globular clusters if the binary is not circularized at merger. If BH are born with modest natal spin, these conclusions will apply even to $2g-mg$ mergers. A $3g$ BH has a mass upper limit, $M_{3g,{\rm max}} \leq 3M_{\rm gap,lower}$ where $M_{\rm gap,lower}$ is the lower bound on the upper mass gap in the natal BH mass distribution \citep[see e.g.][]{FarmerGap, o3a_pop}. For $M_{\rm gap,lower}=40M_{\odot})$, $M_{3g,{\rm max}} \sim 120M_{\odot}(M_{\rm gap,lower}/40M_{\odot})$, would be the maximum mass of a $3g$ progenitor. Expectation $3g$ BH masses could be $\sim 90M_{\odot}$ or less given median masses ($\sim 30M_{\odot}$) observed in mergers, and accounting for energy losses due to GW \citep{o3a_pop}.

If no higher generation mergers are ever detected, it will point strongly towards clusters with shallow potentials (e.g. GCs, open clusters) as the dominant dynamical channel among $1g-1g$ mergers, \textit{and} very low natal spins for black holes. However, given the population spin measurements of \citep{o3a_pop}, it may be challenging to accommodate low natal spins (current measurements suggest $a \sim 0.1-0.2$).

If BH natal spins turn out to be modest (dimensionless spin parameter $a\sim 0.2$) rather than nearly zero, we \emph{require} a large escape velocity along with a high density of BH, to \emph{efficiently} produce dynamically assembled, hierarchical mergers. Then, hierarchical mergers must occur in deep gravitational potential wells, and attempting to distinguish between mergers in active galactic nuclei (AGN) and quiescent galactic nuclei (GN) becomes important.

We proceed as follows: we use the formalism of \cite{McK18} to consider the parameters influencing the rate of BBH mergers in quiescent (gas-poor) and active (gas-rich) galactic nuclei, both containing supermassive black holes (SMBH). We determine the variables governing the relative rates in each environment and show, for any plausible set of nuclei, that AGN will dominate the BBH merger rate relative to gas-poor nuclei containing an SMBH. Finally, we discuss the astrophysical consequences for current and future observers.

\section{Methods}
\label{sec:methods}
We can write a simple but illuminating
`Drake equation' for the rate density of BBH mergers in all galactic nuclei (GN), both active (A) and quiescent (G) as \citep{McK18}:
\begin{equation}
{\cal{R}}_{G}+{\cal{R}}_{A}=\frac{N_{BH}f_{b}n_{GN}}{t_{b}}
\label{eq:R}
\end{equation}
where $R_{A,G}$ is the rate in $\rm{Gpc^{-3} yr^{-1}}$ from each environment, $N_{BH}$ is the number of stellar origin BH per nucleus, $f_{b}$ is their binary fraction, $n_{GN}$ is the number density 
of galactic nuclei $\rm{Gpc^{-3}}$ in the Universe and $t_{b}$ is the average binary lifetime in years. We assume that all galactic nuclei are either active or quiescent, i.e. $n_{GN}=n_{A}+n_{G}$ and we assume that the fraction of galactic nuclei that are active is $f_{AGN}=n_{A}/n_{G}$. The merger rate density is then 
\begin{equation}
{\cal{R}}_{G}+{\cal{R}}_{A}=\left[\frac{N_{G,BH}f_{G,b}(1-f_{AGN})}{t_{G,b}}+\frac{N_{A,BH}f_{A,b} f_{AGN} }{t_{A}}\right]n_{GN}
\label{eq:R_QvsA}
\end{equation}
where $G$ or $A$ modifies the previous quantities for galactic nuclei or active galactic nuclei and $t_{A}$ is a characteristic timescale associated with binaries in AGN. In practice, we can use
\[
t_{A} =
\begin{cases}
\tau_{AGN} & \text{if } t_{A,b}<\tau_{AGN}\\
t_{A,b} & \text{if } t_{A,b}=\tau_{AGN}\\
t_{A,b}/\tau_{AGN} & \text{if } t_{A,b}>\tau_{AGN}
\end{cases}
\]
where $t_{A,b}$, is the average binary lifetime in AGN and $\tau_{AGN}$ is the lifetime of the AGN disk. If $t_{A,b} < \tau_{AGN}$, then most binary mergers happen quickly and early in the lifetime of the AGN, but we must still average the observed rate over the entire AGN lifetime; if $t_{A,b} > \tau_{AGN}$, we will see fewer mergers in AGN, scaled by the ratio of the average binary lifetime to the AGN lifetime. Here we are assuming that all binaries are pre-existing and no new binaries form in the disk (we will alter this assumption later on). Note that we do not divide the AGN population into BH embedded within and without the disk. This is because the existence of the gas disk can harden the distribution of binary semi-major axes in the galactic nucleus, \emph{even for those binaries that do not end up embedded in the AGN disk for their entire orbit} \citep{Tagawa20}. Assuming that $N_{BH}$ is initially similar in both active and quiescent nuclei, and assuming $f_{AGN}$ is small, we write the total BBH merger rate density from galactic nuclei ($R_{G}+R_{A}$) as
\begin{equation}
{\cal{R}}_{G}+{\cal{R}}_{A} \approx N_{BH}n_{GN}\left(\frac{f_{G,b}}{t_{G,b}}+\frac{f_{AGN}f_{A,b}}{t_{A}}\right).
\label{eq:Rapprox}
\end{equation}
So, AGN will dominate the BBH merger rate density from galactic nuclei if the ratio of the rates (${\cal{R}}_{A/G}={\cal{R}}_{A}/{\cal{R}}_{G}$) is
\begin{equation}
{\cal{R}}_{A/G}=f_{AGN}\left(\frac{t_{G,b}}{t_{A}}\right) \left(\frac{f_{A,b}}{f_{G,b}}\right)> 1.
\label{eq:compare}
\end{equation}
Thus, which type of nucleus dominates depends on the fraction of galactic nuclei which are active $f_{AGN}$, the ratio of the binary lifetime in quiescent nuclei to the relevant timescale in active nuclei $t_{G,b}/t_{A}$, and the ratio of the binary fractions in active to quiescent nuclei, $f_{A,b}/f_{G,b}$. The latter ratio should be at least 1, and cannot be larger than $\sim 100$, since: 1) binary fractions are typically driven to larger values by the introduction of a gas disk \citep[e.g.][and others]{McK18, Secunda19, Yang20, Tagawa20}; 2) binary fractions in quiescent nuclei are expected to be $O(0.1-0.01)$ \citep[e.g.][]{AntPer12}; and 3) binary fractions in AGN cannot be larger than 1, and are probably $O(0.1)$, leading to $f_{A,b}/f_{G,b}\sim O(1-10)$.

So, apart from factors roughly of order unity, the ratio of rates is determined by the fraction of galactic nuclei that are active (and for our purposes, `active' refers to nuclei with disks dense enough to substantially alter the dynamics of stars and BH that interact with it), and the ratio of the binary lifetimes in quiescent nuclei to active nuclei. Simulations allow us to infer approximately $t_{A,b}\sim 0.1-1{\rm Myr}$ \citep[e.g.][]{Baruteau11,Tagawa20,Secunda20a,Yang20,McK20b}.
AGN lifetimes are substantially uncertain, but $0.1{\rm Myr} < \tau_{AGN} < 100{\rm Myr}$ \citep[e.g.][]{Schawinski15}.
Average binary times to merger in quiescent nuclei can be estimated them from the rates found by NSC BBH merger models and a rearrangement of equation \ref{eq:R}
\begin{equation}
    t_{G,b}=\frac{N_{BH}f_{G,b}n_{GN}}{{\cal{R}}_{G}}.
\label{eq:tgb}
\end{equation}
While the actual binary lifetime may vary by the type of NSC (cored, cusped, mass segregated; see also below), the rate of mergers from NSCs implies a characteristic average timescale over all quiescent nuclei; we can apply that timescale to evaluate the relative importance of the presence or absence of a gas disk, which serves as a substantial accelerant of BBH mergers. We note that this `average' binary lifetime does not characterize the actual lifetime of an individual binary in a gas-poor NSC, since in the case where a BBH is ionized before it can merge, the binary's time to merger is infinite; indeed, most NSC-triggered BBH mergers happen extremely quickly, or not at all, and the average binary lifetime is a way of comparing the types of nuclei, while accounting for the high rate of ionization events.

\section{Realistic Galactic Nuclei}
 The stellar remnant population in galactic nuclei is expected to consist of some combination of the results of dynamical decay (including that of globular clusters \citep{Generozov18}, dwarf galaxies and minor mergers \citep{Antonini14}), as well as stochastic episodes of star formation (including as a result of AGN activity).  Where the SMBH mass is $M_{\rm SMBH} \leq 10^{7.5}M_{\odot}$, dense nuclear star clusters (NSCs) and the SMBH both seem to both contribute significantly to the central potential \citep{Seth08}. For $M_{\rm SMBH} \geq 10^{7.5}M_{\odot}$ the central potential appears to be dominated by the central SMBH. In this case, nuclear stellar populations should still be present, but are insufficiently massive to dominate the potential. From \S\ref{sec:methods} above, the relative rates from quiescent and active galactic nuclei depends primarily on $t_{G,b}$, and secondarily on $f_{A,b}$ and $f_{G,b}$. Here we elaborate on some of the properties of galactic nuclei on which ($t_{G,b},f_{G,b}$) depend.

\subsection{Binary lifetimes}
The average lifetime of a BBH in an NSC depends on the average mass function, binary fraction ($f_{G,b}$), mass segregation, relaxation rate and encounter rate between binaries and tertiaries (including other binaries). 

Metallicity can also play a significant role in the rate of hierarchical mergers expected from NSCs, with a potentially high rate at low metallicity ($Z \sim 10^{-3}-10^{-4}Z_{\odot}$), but with the rate dropping to zero at $Z \geq 0.01Z_{\odot}$ \citep{Mapelli20}. Metallicities are typically high ($Z>0.1Z_{\odot}$) in galaxies out to redshift $z>3$ with stellar masses $M_{\ast} > 10^{9}M_{\odot}$, reflecting bursts of star formation after gas infall or mergers \citep{Mannucci09}. Likewise, metallicities are typically high in AGN across redshifts out to $z \sim 3$, even becoming super-solar in more massive host galaxies  \citep{Matsuoka18}. Metallicity in present-day globular clusters in our own Galaxy is bi-modal, with peaks at $Z \sim 0.02 Z_{\odot}$ and $Z \sim 0.2Z_{\odot}$ \citep{Muratov10}. So, a combination of AGN activity and star formation in the nucleus (due to major or minor mergers) might be expected to enhance the metallicity of an NSC over the $Z \geq 0.01Z_{\odot}$ threshhold, even with a large population contribution from very low metallicity globular clusters. Certainly, we should expect modest metallicities in NSC at least out to $z \sim 2$, reflecting mergers, infall, star formation and globular cluster arrival via dynamical friction. As present GW detections of BBH mergers are restricted to $z \sim 2$, for the rest of this paper we shall ignore the role of metallicity in establishing overall rate comparisons.

\citet{Antonini16} find that in NSCs without SMBH, the expected rate of mergers is O($10^{2} \rm{Gyr}^{-1} \rm {NSC}^{-1}$). But, the potentially high rate of mergers per nucleus found by \cite{Antonini16} applies only to those lower mass systems missing an SMBH. \cite{2020NeumayerRev} note that dwarf galaxies (stellar masses $<10^{9}M_{\odot}$) are unlikely to host an SMBH, but above that mass threshold, a rising fraction of nuclei do host an SMBH, while also frequently hosting NSCs. Integrating over their expected galaxy mass function, \citet{Antonini16} find an overall merger rate density of $1.5~{\rm Gpc}^{-3}{\rm yr}^{-1}$, which is typically subdominant to other merger channels. Merger hierarchies in SMBH-less NSCs in dwarf galaxies may also occur \citep{FragSilk20}, but here we will focus on galactic nuclei containing SMBH in non-dwarf galaxies.

The merger rate in an NSC in the presence of an SMBH---absent a gas disk---is expected to be quite low \citep{AntPer12}. This is because most binaries within $\sim 0.1R_{\rm inf}$ of the SMBH are expected to be softened by tertiary encounters, where
\begin{equation}
    R_{\rm inf}=\frac{GM_{\rm SMBH}}{\sigma^{2}} \sim 0.1{\rm pc}\left(\frac{M_{\rm SMBH}}{10^{8}M_{\odot}} \right)\left( \frac{\sigma}{200{\rm km/s}}\right)^{-2},
\end{equation}
and $\sigma$ is the typical velocity dispersion in the NSC. Merger rate densities due to Kozai resonances span $~{\rm O}(10^{-3}-10^{-1})~{\rm Gpc}^{-3}{\rm yr}^{-1}$ in NSCs around an SMBH depending on whether an NSC is cored (low rate), cusped, or mass segregated (highest rate) \citep{AntPer12}. This rate density assumes a BH binary fraction of $f_{\rm b} \sim 0.1$ and a BH number density that falls off as $r^{-2}$. We additionally assume $n_{\rm gal} \sim 4 \times 10^{-3}{\rm Mpc}^{-3} = 4 \times 10^{6} {\rm Gpc}^{-3}$ galaxies of Milky Way mass or greater in the local Universe, and each such galaxy has one NSC and one SMBH (or $n_{\rm gal}=n_{\rm GN}$). There may be additional SMBH in galaxies due to minor mergers, but their associated clusters cannot be too massive, otherwise they would be EM-detectable.

The upper end of the rate from \citet{AntPer12}, allows us to deduce from eqn.~\ref{eq:tgb} that 
\begin{equation}
\begin{aligned}
    t_{G,b} = {} & 40 {\rm Gyr} \left(\frac{N_{\rm BH}}{10^{4}} \right)\left(\frac{f_{G,b}}{0.1}\right)
    \left( \frac{n_{\rm GN}}{4 \times 10^{6} {\rm Gpc^{-3}}}\right) \\
    & \left( \frac{{\cal{R}}_{G}}{0.1 \rm{Gpc}^{-3}{\rm yr}^{-1}}\right)^{-1}
\end{aligned}
\end{equation}
where we assume $N_{BH} = 10^{4} {\rm pc}^{-3}$ (see \S\ref{sec:nbh} below). 

By contrast, in AGN, binary ionizations are expected to be rare. This is because of the efficiency of gas drag in shrinking the binary semi-major axis to less than that of the hard-soft boundary. This is true even for binaries ejected from the AGN disk \citep{Tagawa20}, or those which interact with the gas relatively briefly (as for those that pass through the disk on inclined orbits). Thus, the binary lifetime has a notable impact on our next set of important parameters, the binary fraction in each environment.

\subsection{Binary fractions}
The binary fraction in our own Galactic center is poorly constrained. Estimates of the binary fraction are often an extrapolation from the observed binary fraction of massive or low mass stars, incorporating likelihoods of disruptive supernova kicks, convolved with random softening or hardening tertiary encounters for those binaries that survive, or form. Several mechanisms act to suppress the binary fraction in the innermost regions of galactic nuclei and we outline them here. 

First, very close to an SMBH, for a binary of mass $M_{\rm bin}$ and semi-major axis $a_{b}$ there is a binary tidal disruption radius $R_{\rm b,T} \propto a_{\rm b} q^{-1/3}$ where $q=M_{\rm bin}/M_{\rm SMBH}$ at which tidal forcing on the binary exceeds its binding energy \citep{Hills88}. This process is directly analagous to the tidal radius $R_{\rm T} \propto R_{\ast} q^{-1/3}$ around SMBH at which the energy in a raised tide on a star exceeds the binding energy of that star yielding a tidal disruption event (TDE) \citep{Rees88,1989IAUS..136..543P}. Binary disruption close to the SMBH takes the form of partner exchange forming a new binary between $M_{\rm SMBH}$ and $M_{1}$, ejecting $M_{2}$ at hypervelocity and is most probable within $\sim O(10-100)$AU of the SMBH, depending on $M_{\rm SMBH}$ and $a_{\rm b}$ \citep{Hills88}.

Second, further from the SMBH, there is still a radius of influence ($R_{\rm inf}$) where we expect most binaries to be softened by tertiary interactions. That is, the binding energy of the binary is less than the average energy in a tertiary encounter ($E_{\rm b}<M_{\rm bin} \sigma^{2}$) where $\sigma$ is the 1-d velocity dispersion of the NSC stars. Here binaries are ionized on a timescale \citep{BinneyTremaine87}
\begin{equation}
    t_{\rm ion} = \frac{1}{16\pi} \left(\frac{M_{\rm bin}}{M_{\ast}}\right) \frac{\sigma}{\rho_{\ast} a_{\rm bin}} \frac{1}{\rm{ln} \Lambda}
\end{equation}
where $M_{\ast}$ is the typical stellar mass in the NSC, $\rho_{\ast}$ is the central stellar mass density ($M_{\odot}/\rm{pc}^{3}$), $\rm{ln} \Lambda $ is the Coulomb logarithm, with $t_{\rm ion} \sim O(10)$Myr-Gyr for plausible ranges of these parameters.

A combination of sources of binary ionization suggests that primordial binaries will be rapidly ionized, deep in the central galactic nucleus. The only way of circumventing the strong ionization tendency in gas-poor nuclei is via a high rate of binary formation, allied with strong eccentricity pumping, which could drive binary hardening faster than ionization. However, for binaries dynamically formed in AGN disks, typical binary eccentricities are significantly higher than for binaries formed in gas-poor nuclei \citep{Samsing20,Tagawa21a}. Recall that low generation mergers in GCs are expected to be eccentric, while high generation mergers from GCs are expected to be circular \citep[e.g.][]{Carl16MCs}. Thus, observations of eccentric high mass BBH mergers are a key discriminant between AGN-driven hierarchical mergers and other dynamical assembly channels.

\subsection{Number of stellar origin BH}
\label{sec:nbh}
 In general, in quiescent nuclei, the fraction of all the stars in BH depends on the average stellar mass function and the history and degree of mass segregation \citep[e.g.][]{Generozov18}. However, AGN disks may undergo star formation \citep[e.g.][]{Nayakshin05,Yuri07}, which could enhance the number of stellar origin BH in AGN disks \citep{Stone17}. There are also suggestions that the unusual conditions experienced by stars embedded in an AGN disk could further enhance production of stellar origin BH \citep{MatteoAdam20}. In general, from a number of different approaches to the problem, the number of stellar origin BH in our own galactic nucleus is expected to be $\rm{O}(10^4){\rm pc^{-3}}$ \citep{Morris93,Miralda00,Hailey18, Generozov18}.

These processes are sufficiently uncertain that we will not attempt to include them in our considerations below, but we note that there are essentially no suggestions for mechanisms to \emph{suppress} $N_{A,BH}$ relative to $N_{G,BH}$, at the onset of an active period, and we therefore conservatively assume $N_{A,BH}/N_{G,BH}=1$ throughout.

\begin{figure}
\includegraphics[width=\linewidth]{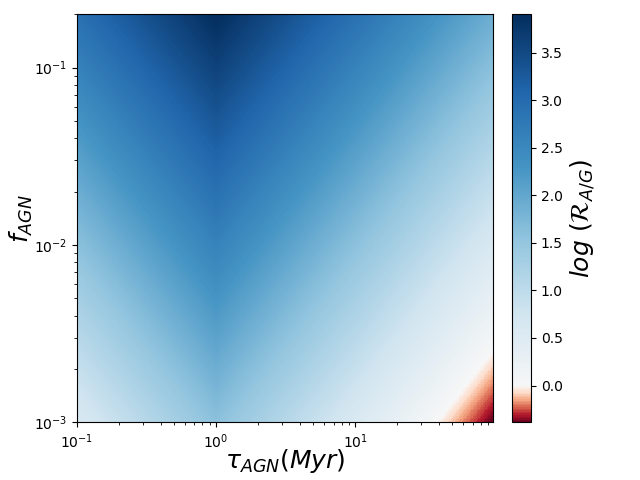}
\caption{Relative rate densities of BBH mergers in active versus quiescent galactic nuclei. For nuclei containing an SMBH, assuming $f_{A,b}/f_{G,b} = 1$, $t_{G,b}=40 {\rm Gyr}$, $t_{A,b}=1 {\rm Myr}$, we find that almost regardless of the fraction of galactic nuclei that are involved in accelerating BBH mergers ($f_{\rm AGN}$) or AGN disk lifetime ($\tau_{\rm AGN}$), AGN dominate the rate of BBH mergers from deep potential wells (blue). For very small $f_{\rm AGN}$ and very large $\tau_{\rm AGN}$, it is possible for quiescent nuclei to make a substantial contribution (red, see text for more detail); however, if we assume a more realistic $f_{A,b}/f_{G,b} = 10$, there is no region of parameter space where AGN are not dominant.
\label{fig:relrates}}
\end{figure}

\section{Results}
Using the estimates above, we can parameterize eqn.~\ref{eq:compare} as
\begin{equation}
{\cal{R}}_{A/G} \sim 400 \left(\frac{f_{AGN}}{0.01}\right) \left(\frac{t_{G,b}/t_{A}}{40{\rm Gyr}/1{\rm Myr}}\right) \left(\frac{f_{A,b}/f_{G,b}}{1}\right).
\label{eq:param_compare}
\end{equation}
Note that ${\cal{R}}_{A/G}$ could be one or two orders of magnitude larger than in eqn.~(\ref{eq:param_compare}) if $t_{A} \sim O(0.1){\rm Myr}$ and if $f_{A,b}/f_{G,b} \sim 10$.

Figure~\ref{fig:relrates} shows the ratio of merger rate density in AGN to merger rate density in GN (${\cal{R}}_{A/G}$) as a function of AGN lifetime ($\tau_{\rm AGN}$) and the fraction of galactic nuclei that are active ($f_{\rm AGN}$). Since we assume the typical time to merger in an AGN disk is $1$Myr, the highest rate enhancement of AGN/GN occurs for $\tau_{\rm AGN}=1$Myr. The figure conservatively assumes $f_{A,b}/f_{G,b} = 1$. We can easily see that the only region of parameter space where quiescent nuclei are competitive with or surpass active nuclei as contributors to the BBH merger rate is where AGN are extremely long lived \textit{and} if only the rarest of nuclei ($f_{\rm AGN} \ll 0.01 $) have disks which can act as BBH merger accelerators. If we more realistically assume $f_{A,b}/f_{G,b} = 10$, we will \textit{always} find ${\cal{R}}_{A/G}>1$.

Thus, if any $ng-mg$ mergers are observed at $z \leq 2$, where $n>3$, they must have originated in an AGN. Further, if natal BH spins are shown to be non-negligible, at $z \leq 2$, $ng-mg$ mergers with $n>2$ must also have originated in AGN. Finally, if natal BH spins are non-negligible, eccentric mergers with $n>1$ (i.e. all eccentric hierarchical mergers) must originate in AGN.
Our results depend almost entirely on the very large escape velocity of galactic nuclei with an SMBH (compared to that of GCs), coupled with the enormous difference in average binary lifetime between active and quiescent nuclei. There is a secondary dependence on the binary fractions in each environment, but given other evidence pointing towards relatively short-lived AGN episodes ($\tau_{\rm AGN} \ll 40$Myr, \citet{Shen21}), even this effect is likely to be irrelevant. We further note that this result does not depend on any assumption of star formation or enhanced BH formation in active over quiescent nuclei. If nuclear activity leads to enhanced BH production, AGN become still more important as locations for hierarchical mergers. One caveat to the reasoning above is that if most AGN episodes are very short lived ($<0.1$Myr), then the AGN disk in these cases may only be the catalyst for the production of a population of very hard BH binaries that then go on to merge via tertiary encounters post-AGN. Multiple short-lived AGN episodes would still allow for multiple such phases of AGN BBH catalysis.

\section{Consequences for AGN}
Here we discuss some of the implications of the arguments above and outline some observational tests that might be performed to measure the contribution of the AGN channel to GW detected BBH mergers. We can determine the fraction of BBH mergers from the AGN channel ($f_{\rm BBH,AGN}$) by identifying the rarest (especially hierarchical) events they uniquely produce. Measuring the rate of those rare events, we can then use models to determine the fraction of remaining (non-hierarchical) mergers that must also come from the AGN channel, and what fraction must come from other channels. We also consider what AGN astrophysics we can learn (with caveats) from GW and multimessenger observations.

\subsection{Clues for events unique to AGN}
There are a handful of clues we can search for among GW-observed BBH merger events that indicate an origin in a deep potential well, which therefore must be from an AGN. Among these clues are: IMBH formation events, significantly asymmetric mass ratios with a very large primary mass, and eccentric mergers. Additionally, very asymmetric mass ratio mergers (at any mass) are signatures of gas processes unique to AGN.

LIGO-Virgo is beginning to detect IMBH formation events, e.g. GW190521 \citep{GW190521PRL,GW190521}. This event is exceptional in many respects: the total mass is $>100M_{\odot}$ (an IMBH); the mass of both progenitors is $>50_{\odot}$  i.e. in the pair instability mass gap \citep[although see][]{Fishbach20}; both component spins were not small and aligned; and there is some evidence for non-zero eccentricity, though this may be degenerate with spin misalignment \citep{RomeroShaw20}. All of these characteristics point to a dynamical process of assembly \citep[e.g.][]{Tagawa21a,Zrake20,Samsing20}\footnote{A candidate EM counterpart was also reported in an AGN \citep{Graham20a}; if the association is correct, it clearly lends further strength to the arguments for dynamical assembly, however our arguments do not rest on the association.}. 
We note that though initial findings of the likelihood of GW190521 being a hierarchical merger were ambiguous \citep{GW190521,Fishbach20}, the ambiguity rests on the prior expectation of the relative rates of hierarchical mergers to $1g-1g$ mergers. For a sufficiently strong prior, a Bayesian parameter estimation will be forced to find the region of permitted parameter space that agrees with both the data and the prior being enforced. In the AGN channel, hierarchical mergers with progenitors in the mass gap are sufficiently common \citep{McK20b} that, if the relative rates from the AGN channel were the enforced prior, GW190521 would have 2 progenitors in the upper mass gap, as the single-source parameter estimation implies.

If the parameters of GW190521 are as described in \cite{GW190521PRL,GW190521}, a hierarchical merger scenario is the most likely origin. The maximum mass of most $1g$ black holes could be as low as $M_{\rm gap, lower} \sim 35 M_{\odot}$---\citep[see e.g. Fig 16 in][]{o3a_pop}, which suggests 2g merged BH are $<70M_{\odot}$ in mass. Thus GW190521 ($85M_{\odot}+66M_{\odot}$) could have been a $3g-2g$ merger. If black hole natal spins are $a \sim 0.2$ \citep[consistent with][]{o3a_pop}, such a high generation merger must have formed in a deep potential well, and by our arguments above, it \textit{must} have come from an AGN disk. If the merger was $3g-2g$ and eccentric \citep[as argued by][]{RomeroShaw20}, it also \textit{must} have come from an AGN disk, regardless of natal spins.

There are additional suggestions of hierarchical mergers in the literature, though none as strong as GW190521. Nevertheless, we note that GW170729 \citep{O2}, GW190412 \citep{GW190412}, and GW190814 \citep{GW190814} have been specifically considered candidate hierarchical mergers \citep[and see][for an excellent review]{GerosaFishbach21}. The latter two events are also notable as candidate AGN-driven mergers, irrespective of their generational status, due to their unequal mass ratios (see more below). Besides these, there are 5 additional events in \cite{gwtc2} with primary masses likely $>50 M_{\odot}$, again, making them high-probability hierarchical candidates. Most recently, GW190426$\_$190642 ($M_{\rm BBH} \sim 184 M_{\odot}$) and GW190403$\_$051519 ($M_{1} \sim 88 M_{\odot}$) are extremely strong candidates for hierarchical mergers, if they are astrophysical \citep{gwtc21}. We can expect to see additional candidates from the upcoming release of the full LIGO-Virgo O3 catalog\footnote{Apart from population considerations, a combination of the publicly released sky area and source distance gives a rough estimate of the chirp mass of a GW event; if astrophysical, S200208q is very likely to be more massive than GW190521.}.

Modestly asymmetric mass ratio mergers, especially those at 1:2 or 1:3 are more likely to be hierarchical mergers (and thus dynamical mergers), since these ratios are the result of integer combinations from some base population, but these can be produced by GCs under particular circumstances \citep{Carl21}. However, the most asymmetric mass ratio mergers (1:10 and more extreme), are only likely to form in an AGN-driven merger environment. This is because, in a gas-poor dynamical environment, exchange interactions tend to sort binaries towards equal mass, though 1:2 and 1:3 events occur occasionally, especially between the most massive object in the cluster and a less massive $1g$ partner. By contrast, gas disks produce mass-dependent migration torques in AGN, which naturally produces asymmetric mass ratio mergers \citep[see e.g.][though these always remain a minority of all BBH mergers]{Secunda20b,McK20b,Yang20,Tagawa21a}. In addition, very extreme mass ratio mergers are uniquely produced if AGN disks harbor migration traps \citep{Bellovary16}, which allow the growth of very large ($>500M_{\odot}$) IMBH \citep{McK20a,McK20b}.

\subsection{AGN fraction of total BBH merger rate}
The currently measured BBH merger rate density is ${\cal{R}}_{\rm BBH} \sim 24^{+15}_{-9}\rm{Gpc^{-3}~yr^{-1}}$\citep{o3a_pop}. 
Including a full accounting of uncertainties yields a BBH merger rate from AGN of ${\cal{R}}_{\rm AGN} \sim 10^4-10^{-4}~\rm{Gpc^{-3}~yr^{-1}}$ for a priori equally valid parameter choices for AGN disks and NSCs \citep{McK18}. Tighter parameter ranges (${\cal{R}}_{\rm AGN} \sim 0.1-60 \rm{Gpc^{-3}~yr^{-1}}$) have been presented \citep{Grobner20, Tagawa20c}, however these do not account for the possibility of multiple AGN episodes in the same nucleus, but do assume more realistic upper limits on the number of BH in a nucleus.
 
GW190521 is the event most likely to have happened in an AGN. If it did, there are two possible locations: 1) at a migration trap, or 2) elsewhere in the disk (which we will call the `bulk'). \citet{McK20b} find $O(15)$ mergers at a trap/Myr (for a 1Myr AGN lifetime and a large radius disk) which implies ${\cal{R}}_{\rm trap} \sim 1 {\rm merger}/70{\rm kyrs}$ per trap per AGN. If we say that only quasars or the most luminous Seyfert AGN are responsible for AGN channel BBH mergers, then $n_{\rm AGN} \sim 0.01n_{\rm GN}$ of all galaxies ($f_{\rm AGN}=0.01$), or $n_{\rm AGN} \sim 4 \times 10^{4}$ AGN/$\rm{Gpc}^{3}$. If we simply assume each such AGN has a trap, then the overall trap merger rate density is ${\cal{R}}_{\rm trap} n_{\rm AGN} \sim 0.6 {\rm Gpc^{-3} yr^{-1}}$, which is approximately the upper limit to the rate of IMBH formation mergers seen by LIGO \citep{GW190521}. Assuming a $10:1$ ratio of mergers in the bulk disk to the trap \citep{McK20b}, we find ${\cal{R}}_{A} \sim 6 {\rm Gpc^{-3} yr^{-1}}$, or $f_{\rm AGN,BBH} \sim 0.25$. If the AGN disk is radially smaller than assumed in \cite{McK20b}, then the ratio of bulk to trap mergers decreases, and $f_{\rm AGN,BBH} < 0.25$.
If instead the merger happened in the bulk disk, i.e. away from a trap, or traps do not exist, then the hierarchical nature of GW190521 corresponds to a $\sim 1/20$ bulk merger event \citep{McK20b} and ${\cal{R}}_{\rm A}$ could be as high as O($20{\rm Gpc}^{-3}{\rm yr}^{-1}$), or $f_{\rm AGN,BBH} \sim 0.8$.

Interestingly, GW190521 is not the only event in \cite{gwtc2} that points to a notable $f_{BBH,AGN}$. GW190814 is a $q=M_{2}/M_{1} \sim 0.1$ merger that is either a BH-BH or a BH-NS merger \citep{GW190814}. Interestingly, $q=0.1$ is the expectation value for $q$ of a BH-NS merger in an AGN disk \citep{McK20b}. For BBH mergers in the bulk AGN disk, a $q=0.1$ merger is a $\sim 5\%$ occurrence \citep{McK20b}, again implying an AGN-driven merger rate $\sim 20{\rm Gpc}^{-3}{\rm yr}^{-1}$.

If the fraction of BBH merger events that come from the AGN channel is relatively large ($f_{\rm BBH, AGN} \geq 0.25$), this implies the AGN in which BBH mergers are occurring are relatively short lived ($<5$Myr), and relatively dense ($\rho>10^{-11}{\rm g/cm^{-3}}$). The short AGN lifetimes are required by the observed $\chi_{\rm eff}$ distribution \citep{o3a_pop}; if most mergers were originating in long-lived AGN, gas accretion would have aligned BH spins with the angular momentum of the gas disk. But that would produce spins also aligned (or anti-aligned) with the orbital angular momentum of the binary, thus producing more extreme values of $\chi_{\rm eff}$ \citep{McK20b}. However, in order to rapidly merge black holes in shorter-lived disks, gas capture of inclined orbiters must be efficient, and that requires high gas densities \citep{Fabj20}.

Longer-lived ($\geq 5$Myr) or low density ($\rho < 10^{-11}{\rm g/cm^{-3}}$) AGN disk can certainly exist; they simply cannot substantially contribute to the measured ${\cal R}_{\rm BBH}$. We should therefore take care in generalizing our inferences of AGN properties from GW detections of BBH mergers, since such detections are biased towards BBH mergers in dense, shorter-lived AGN disks.

From a multi-messenger perspective, this is mixed news---dense, short-lived disks should also be more the more luminous ones. This makes searching for direct EM counterparts harder (the AGN is brighter) \citep{McK19a,Graham20}, but also means there should be fewer such luminous AGN in each LIGO-Virgo error volume. This means it will be easier to use indirect statistical inference methods \citep[e.g.][]{Bartos17} to determine $f_{\rm AGN,BBH}$ from GW observations and archival AGN catalogs alone---provided such catalogs have adequate completeness and reliability out to the LIGO-Virgo horizon \citep{FordWP}.

Looking forward, LISA \citep{LISA} could detect a large population of IMBH-SMBH mergers, if IMBH are formed at migration traps in AGN at high efficiency. LISA can also detect IMRIs which should also occur at migration traps when an IMBH merges with a lower mass BH delivered by gas torques. \cite{McK20b} find that the median mass at the migration trap is $\sim 150M_{\odot}$ within 1Myr and the mass at the trap grows as $\tau_{\rm AGN}^{3/2}$. If AGN are relatively short-lived ($\sim 1$Myr), the IMBH will not grow much beyond a few hundred $M_{\odot}$. If some AGN are longer lived, a $\sim 10^{3}M_{\odot}$ IMBH could build up; however, these AGN will be different from the dominant source of BBH mergers seen by ground-based GW detectors.

Finally, we note that the ratio of hierarchical mergers to $1g-1g$ mergers varies by dynamical channel, and is largest for mergers in AGN. Measuring this ratio can help constrain the branching fraction between channels, and especially between dynamical channels. If high generation hierarchical mergers are found to be sufficiently common, it will help constrain the branching fraction between GC and AGN. This represents a critical tool for using the GW-measured distribution functions of such parameters as mass and spin to constrain \textit{multiple channels}. If a given model channel produces known mass, spin, eccentricity, etc. distributions, and that model's branching fraction can be measured using rare or unique events, we can subtract the distribution of that channel from the overall observed parameter distributions. The residual distribution will then represent only the remaining channels---and if it is possible to do this sequentially, as might be achieved for multiple dynamical channels, we can better use GW observations to constrain important unknown physical processes, such as  common envelope processes in isolated binary evolution.

\section{Conclusions}
Binary black hole (BBH) merger rates should always be higher in active galactic nuclei than in quiescent galactic nuclei, if those nuclei contain an SMBH. This is primarily due to the difference in the average time to merger for a BBH in each environment, which in turn is driven by the high ionization rate of binaries in gas-poor NSCs. There is a secondary effect due to the enhanced rate of binary formation in AGN; the formation of BH in AGN is likely a small additional factor. Because high-generation hierarchical mergers are unique signatures of dynamical processes in a deep potential well (regions with a large escape velocity), we can use any unambiguous detection of such a merger as a probe of the overall merger rate of BBH from AGN.

In particular, if natal BH spins are typically near zero, AGN must be uniquely responsible for $n$th generation mergers where $n>3$, regardless of the properties of the merger. If natal BH spins are modest ($a \sim 0.2$) \textit{or} a merger is eccentric, AGN must be uniquely responsible for mergers where $n>2$.
Finally, if natal BH spins are modest \textit{and} a merger is eccentric, AGN must be uniquely responsible for mergers where $n>1$. Since the maximum mass of the initial black hole mass distribution remains uncertain, and progenitor spin measurements are frequently also uncertain, it is still difficult in most cases to distinguish which generation a particular merger may be, and we are hopeful that future observations will clarify the underlying distributions of masses and spins for $1g-1g$ mergers.

Current observations of candidate hierarchical mergers imply that $f_{\rm AGN,BBH} \sim 0.25-0.8$, the fraction of all BBH mergers that could be accounted for by AGN. In principle, we can use such a measurement to constrain the mass, spin, etc. distributions from this and from other channels. We note that the AGN contributing the most to the BBH merger rate will be those with shorter lifetimes ($<5$Myr) and larger gas densities ($\rho > 10^{-11}~{\rm g~cm^{-3}}$). These short lifetimes make it difficult to build up very large IMBH masses ($\sim 10^{3}M_{\odot}$), which will limit the rate of formation of substantial IMBH-SMBH binaries easily detectable by LISA. Longer-lived AGN disks may still exist, and they may still produce some substantial IMBH-SMBH binaries; however, these must be \textit{different} from the AGN disks producing BBH merger detections. 

We therefore should be cautious in our inferences about the properties of AGN disks from GW observations. BBH mergers likely probe only the most luminous AGN; other types of AGN do exist, and their parameters (lifetimes, densities, volume filling factors) will need to be probed via other methods (possibly including GW observations of IMBH-SMBH binaries using LISA).

{\section{Acknowledgements.}}
KESF \& BM are supported by NSF AST-1831415 and Simons Foundation Grant 533845 and by the Center for Computational Astrophysics of the Flatiron Institute.
KESF wishes to thank Nathan Leigh, Ari Maller, Carl Rodgriguez, and David Zurek for helpful discussions on cluster modelling and evolution. KESF \& BM wish to thank Tom Callister, Will Farr, and Eric Thrane for helpful discussions on BBH populations, BBH spins, and data visualization. We also wish to thank the many attendees of the Gravitational Waves Group Meeting at CCA, especially Floor Broekgarden, for motivating many implications of this work.
Finally, we would like to thank all of the essential workers who put their health at risk during the COVID-19 pandemic, without whom we would not have been able to complete this work.

\section*{Data Availability}

Data and code used in the creation of Figure 1 is available at https://github.com/saavikford/AGN-relative-rates.
Any other data used in this analysis are available on reasonable request from the first author (KESF).
 



\bibliographystyle{mnras}
\bibliography{refs_thermo} 








\bsp	
\label{lastpage}
\end{document}